\documentclass[conference]{IEEEtran}
\IEEEoverridecommandlockouts

\usepackage{amsmath,amssymb,amsfonts}
\usepackage{algorithmic}
\usepackage{graphicx}
\usepackage{textcomp}
\usepackage{xcolor}
\usepackage[hyphens]{url}
\usepackage{hyperref}
\hypersetup{colorlinks,allcolors=blue}
\usepackage[backend=biber, style=ieee]{biblatex}
\usepackage{float}
\usepackage{balance}
\usepackage{tabularx}
\usepackage{orcidlink}

\def\BibTeX{{\rm B\kern-.05em{\sc i\kern-.025em b}\kern-.08em
    T\kern-.1667em\lower.7ex\hbox{E}\kern-.125emX}}
\begin{document}

\title{Securing Agentic AI Systems - A Multilayer Security Framework\\
}

\author{\IEEEauthorblockN{Sunil Arora\,\orcidlink{0009-0007-3066-3461}, John Hastings\,\orcidlink{0000-0003-0871-3622}}
  \IEEEauthorblockA{
    \textit{The Beacom College of Computer \& Cyber Sciences} \\
    \textit{Dakota State University, USA}\\
    sunil.arora@trojans.dsu.edu, john.hastings@dsu.edu \\
    }
}

\maketitle

\begin{abstract}
Securing Agentic Artificial Intelligence (AI) systems requires addressing the complex cyber risks introduced by autonomous, decision-making, and adaptive behaviors. Agentic AI systems are increasingly deployed across industries, organizations, and critical sectors such as cybersecurity, finance, and healthcare. However, their autonomy introduces unique security challenges, including unauthorized actions, adversarial manipulation, and dynamic environmental interactions. Existing AI security frameworks do not adequately address these challenges or the unique nuances of agentic AI. This research develops a lifecycle-aware security framework specifically designed for agentic AI systems using the Design Science Research (DSR) methodology. The paper introduces MAAIS, an agentic security framework, and the agentic AI CIAA (Confidentiality, Integrity, Availability, and Accountability) concept. MAAIS integrates multiple defense layers to maintain CIAA across the AI lifecycle. Framework validation is conducted by mapping with the established MITRE ATLAS (Adversarial Threat Landscape for Artificial-Intelligence Systems) AI tactics. The study contributes a structured, standardized, and framework-based approach for the secure deployment and governance of agentic AI in enterprise environments. This framework is intended for enterprise CISOs, security, AI platform, and engineering teams and offers a detailed step-by-step approach to securing agentic AI workloads. 
\end{abstract}

\begin{IEEEkeywords}
Agentic Artificial Intelligence, AI Security, AI Risks, Secure AI Framework, Lifecycle Security
\end{IEEEkeywords}

\section{Introduction}
Artificial intelligence (AI) enables machines to perceive, reason, learn, and decide \cite{russell2020aima}. Agentic AI is the latest development in the evolution of intelligent systems. Agentic AI systems can make decisions, plan actions, select tools to achieve an outcome, and adjust to changing environments without continuous human control \cite{1acharya, HOSSEINI2025100399}. Their design allows them to pursue defined objectives while responding to new information and conditions in real time. 

Unlike traditional machine learning, AI, or generative models that operate within fixed parameters, agentic AI systems show continuous improvements, reasoning, and autonomous behavior across diverse contexts. This ability of AI agents to achieve autonomy, automated workflows, and decision-making has generated significant interest from industries such as cybersecurity, finance, healthcare, transportation, medicine, and industrial automation \cite{tayiba, PowellmedicineAgenticAI, KARUNANAYAKE202573}, where autonomous operations offer significant efficiency and flexibility.

The same autonomy that makes agentic AI valuable also creates new security concerns. Existing AI security frameworks, such as NIST AIRMF \cite{nist2023ai_rmf}, ENISA \cite{ENISA_2023multilayer}, EU AI Act \cite{Eu2024AIAct}, and ISO/IEC 42001:2023~\cite{ISO42001}, are designed for general and generative AI models and systems that do not account for continuous, autonomous decision-making capabilities. Industry standards and risk frameworks, such as EY Agentic AI risk frameworks \cite{godhrawala}, outline risks but lack the technical guidance and capabilities to mitigate agentic AI risks.  Agentic AI introduces risks, such as unauthorized actions, adversarial interference, goal drift, and data misuse, arising from dynamic interactions with complex environments. In addition, evolving reasoning processes and unpredictable adaptations make it challenging to verify system security, integrity, and compliance once deployed.

As these systems move into high-stakes, regulated domains, the absence of a security framework that considers their full lifecycle development, deployment, operation, and governance has become a critical gap. Organizations face increasing exposure to operational, ethical, and compliance risks if such systems operate beyond their intended boundaries or are exploited by adversaries.

This research addresses that gap by proposing a security framework specifically designed for agentic AI systems. The framework adopts a lifecycle perspective, integrating protection mechanisms from data handling and model integrity to agent control, monitoring, and governance. It aims to help organizations manage confidentiality, integrity, availability, and accountability throughout the system’s operational lifespan, providing an end-to-end standard for secure and reliable use of agentic AI in enterprise environments.

\section{Background}

In 2024, the market value of agentic AI reached USD 5.1 billion. According to Capgemini, it is expected to exceed 47 billion U.S. dollars, with a compound annual growth rate of over 44\% \cite{vailshery_2025_global_agentic_ai_market_value}. This impressive growth highlights the potential of agentic AI to revolutionize industries through autonomous actions and decision-making. As per another survey, in 2024, less than 1\% of enterprise software included agentic AI. By 2028, nearly a third are expected to incorporate agentic AI, greatly enhancing decision-making autonomy \cite{vailshery_2025_global_share_ai}.

AI has evolved significantly since its origins in 1950, when Alan Turing and John McCarthy laid the groundwork for machine learning. It has advanced through many improvements that have shaped its capabilities and applications \cite {angel2022, turing1980}. Initially, it was developed as a rule-based expert system. AI has reached a stage where it can make autonomous decisions, learn from new datasets, and perform complex multistep tasks with minimal or no human oversight and intervention. This section provides an overview of AI’s historical development, its different types and generations, and the emergence of agentic AI, a novel advancement poised to redefine the field. Agentic AI presents new opportunities for automation and efficiency in many fields, but also raises critical security concerns that shall be addressed for its secure and ethical deployment.

AI research noticed a shift in statistical and machine learning methods in the 1980s and 1990s. It enabled computers to learn patterns from data instead of relying on a predefined set of rules. This transition marked the rise of Artificial Narrow Intelligence (ANI), in which AI models performed excellently at specific tasks such as speech recognition and image processing. However, it was still far from broader cognitive abilities. The advancement of deep learning and neural networks in the 2010s revolutionized AI, significantly improving performance in fields such as natural language processing (NLP), autonomous systems, and robotics. However, despite these advancements, traditional AI models still require extensive human oversight \cite{eran, Sayles2024}.

AI systems are generally grouped into three categories: ANI, Artificial General Intelligence (AGI), and Artificial Superintelligence (ASI).

\textbf{ANI} refers to systems designed to perform specific tasks, such as virtual assistants, recommendation engines, or autonomous vehicles \cite{Ruchira}. These systems operate within defined parameters and cannot apply their knowledge outside their programmed scope.

\textbf{AGI} represents a theoretical stage of AI capabilities that can perform any task a human can. It would be able to reason, learn, and apply knowledge across different domains \cite{goertzel2014artificial}. Although AGI has not yet been achieved, ongoing research continues to advance toward this goal.

\textbf{ASI} is the most advanced form that can surpass human intelligence in every area, including creativity, reasoning, and social understanding \cite{barrett2017model}. Its potential development raises ethical, safety, and security concerns about control and societal impact.

Agentic AI refers to the next phase in the evolution of AI, in which systems that demonstrate independence, pursue specific goals, and adapt to given situations \cite{kostopoulos}. Unlike traditional or generative AI models that require detailed prompts for each task, agentic AI has capabilities to determine and perform actions to accomplish a given objective \cite{watson2024guidelines}. These systems can interact with other AI agents, applications, APIs, and physical environments, plan multi-step tasks, and adjust their approach based on feedback and changing conditions \cite{shavit2023practices, Wooldridge_Jennings_1995, Rafael}.

A key distinction between agentic AI and previous AI models lies in their ability to take initiative. For example, a generative AI model like ChatGPT \cite{Fui-Hoon} answers the questions and queries or generates new content, but does not act on its own. Whereas an agentic AI system can take charge or manage tasks, such as booking travel accommodations, scheduling meetings, or fixing software or coding issues, without the need for continuous human inputs. Agentic AI systems leverage language models, reinforcement learning, and multimodal capabilities to operate efficiently in complex scenarios \cite{Dorri}. 

An important advantage of an agentic AI system is its ability to perform tasks beyond what humans can handle. These systems can handle intricate workflows, perform a high number of tasks at a rapid pace, and maintain accuracy and efficiency in critical processes. Organizations can improve productivity and reduce error rates in their products and services by minimizing human involvement in redundant processes \cite{whiting2024rise,mohapatra2019, agentic2025new}. However, agentic AI can introduce challenges such as governance concerns, misalignment with human values, accountability, and security issues. These must be addressed for safe and secure use of agentic AI \cite{garvey2024agentic, ghose2024next, Cohen}. The autonomy and self-directed behavior of these agentic AI systems mean that errors or unintended behaviors can lead to real-world consequences. For example, an agentic AI assistant handling a user’s accounts and finances might perform unauthorized or unintended transactions by mistake, resulting in monetary loss. Similarly, poorly configured AI agents may get compromised by bad actors, resulting in unintended outcomes. It highlights the need for robust AI oversight mechanisms. Cybersecurity vulnerabilities in agentic AI could negatively impact their outcome, business operations, and customers. Developing and implementing security frameworks that include continuous monitoring, user verification, and ethical AI guidelines is essential to mitigate these risks.

\section{Methodology}

This study uses the Design Science Research (DSR) methodology to develop a lifecycle-based security framework for agentic AI systems. DSR is suitable for this work because it emphasizes the creation of practical artifacts grounded in real problems, established knowledge, and systematic evaluation \cite{peffers2007design}. The research process follows core DSR activities: problem identification, objective definition, artifact design, and evaluation.

The primary qualitative method used in this study is a systematic literature review (SLR). The SLR was conducted to identify existing knowledge on AI security, autonomous agent behavior, adversarial risks, and lifecycle governance. Sources and research papers directly related to AI security, autonomous decision-making, agentic AI, or known AI adversarial behaviors were included, ensuring the review remained focused and evidence-driven. The systematic literature review used search terms, including ‘AI security,’ ‘agentic AI,’ ‘AI agents,’ and ‘agentic AI risks.'  To include the most recent developments in AI security, agentic AI, and autonomous agent research, studies, and standards published from 2022 onward were preferred for the review. The findings from the SLR informed both the structure and content of the proposed framework.

For evaluation, the framework security layers and components were systematically mapped to threat categories and adversarial tactics in the MITRE ATLAS to assess the extent to which the proposed controls address known risks in AI systems. This evaluation approach provides framework validity against the industry standard for AI threats and risks. This ensures that the framework is built on existing research, aligns with recognized standards, and can be applied to real-world applications of agentic AI.

\section{Multilayer Agentic AI Security (MAAIS) Framework}

The proposed MAAIS Framework is built on a layered architecture to achieve defense-in-depth and a zero-trust approach. This multi-layered defense model is designed to ensure the secure deployment and operation of agentic AI systems. This architecture provides a defense-in-depth approach to agentic AI security, ensuring that critical security aspects of AI operations, from data protection to user \& access management, are protected against emerging threats.

\subsection{CIAA Model}

The CIA triad in cybersecurity is a foundational model comprising Confidentiality, Integrity, and Availability. This model outlines the guidelines for creating and maintaining security policies that protect data. It ensures that information is protected and is only accessible to authorized users, data is accurate and trustworthy, and systems and data are available when needed \cite{samonas2014cia}. 

\begin{figure}[h]
    \centering
    \includegraphics[width=1\linewidth]{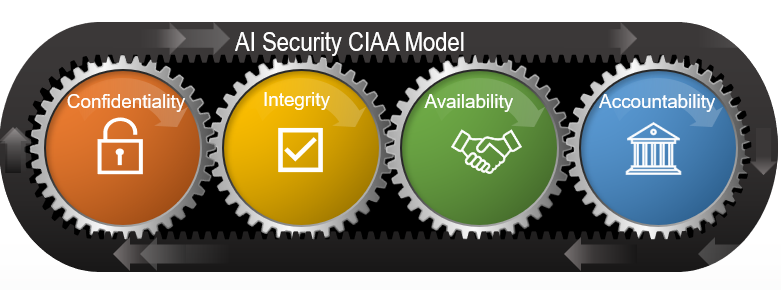}
    \caption{CIAA Model }
    \label{fig:CIAA}
\end{figure}

In agentic AI systems, where decisions and actions may occur without direct human oversight, maintaining accountability is essential for security, transparency, and governance. Accountability enables organizations to identify who is responsible for the AI Agent's decisions and outcomes. Accountability models may vary depending on the organization's and institution's needs, such as when an employee uses personal AI agents for their work or when an application owner uses an AI agent in an application to achieve a business outcome. Accountability prevents misuse and builds trust in AI-driven processes. Hence, it drives, enhances, and enforces explainability, trustworthiness, and transparency. It is a fundamental component of agentic AI security and governance. The augmented CIAA model (Confidentiality, Integrity, Availability, and Accountability), as shown in Fig. \ref{fig:CIAA} provides a comprehensive foundation for securing agentic AI systems throughout their lifecycle.
\subsection{ MAAIS - Agentic AI Security Framework}

\begin{figure}
    \centering
    \includegraphics[width=0.75\linewidth]{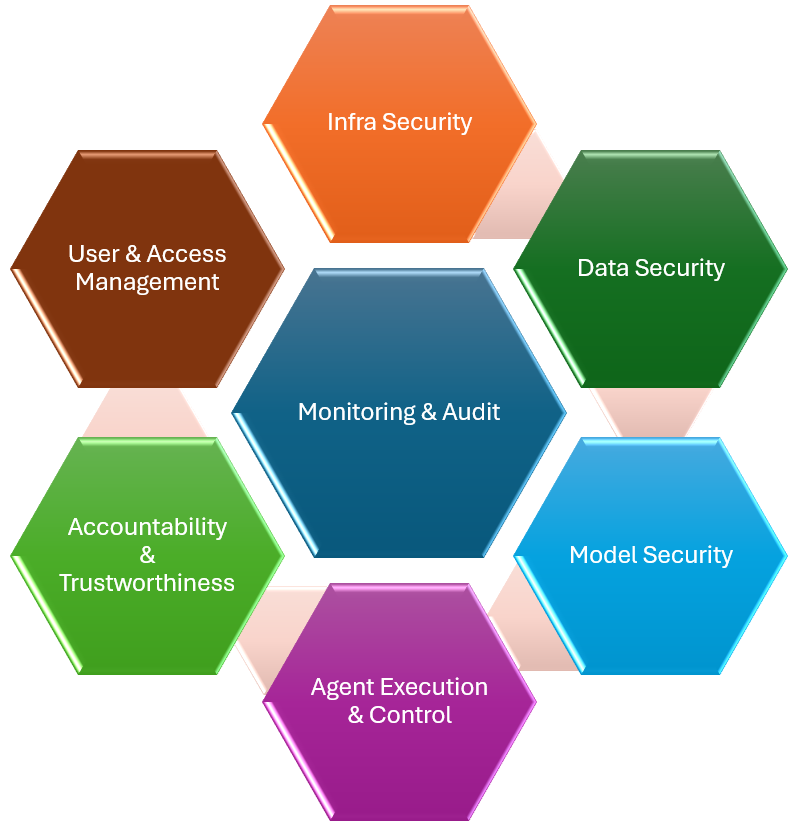}
    \caption{MAAIS – A Seven-Layered Agentic AI Security Framework}
    \label{fig:AgenticAI}
\end{figure}

The MAAIS consists of seven interdependent layers each addressing specific security concerns while collectively ensuring confidentiality, integrity, availability, and accountability (CIAA) in agentic AI deployments (as shown in Fig. \ref{fig:CIAA}). The seven layers (shown in Fig. \ref{fig:AgenticAI}) are detailed in the following subsections and include:
\begin{enumerate}
    \item Infrastructure Security
    \item Data Security
    \item Model Security
    \item Agent Execution \& Control
    \item Accountability \& Trustworthiness
    \item User \& Access Management
    \item Monitoring \& Audit
\end{enumerate}

\subsubsection{Infrastructure (Infra) Layer}
The Infrastructure Security layer is a fundamental building block for the development and operation of agentic AI systems. These systems require a secure and resilient infrastructure that can support real-time interactions, decision-making, various integrations, and cross-environment orchestration. This layer addresses the security of compute, storage, network, hardware, virtualization, and orchestration components across both cloud and on-premise environments.

The infrastructure security covers the following controls:
\begin{itemize}
    \item Secure Hardware
    \item  Secure Compute Environment
    \item Secure Orchestration
    \item Supply Chain Security
    \item Storage Protection
    \item Network Segmentation
    \item Secure Deployment (CI/CD)
\end{itemize}

\subsubsection{Data Security}

The Data Security Layer provides the principal protections that ensure agentic systems operate on trustworthy inputs and that their outputs do not expose sensitive information. Because agentic AI both consumes and (often) persistently stores large amounts of heterogeneous data training corpora, episodic interaction logs, knowledge caches, and user-context memory. This layer addresses confidentiality, integrity, provenance, and regulated use of data across the full lifecycle (collection → preprocessing → training/fine-tuning → inference → retention/deletion). The layer’s design therefore combines cryptographic protections, access governance, provenance and lineage tracking, adversarial-poisoning defenses, privacy-preservation methods, secure orchestration of data pipelines, and auditable retention/erasure policies. The section below outlines the technical controls and how they map to specific attacks and mitigations relevant to agentic AI. 

The data security layer covers the following controls:
\begin{itemize}
    \item Data Confidentiality
    \item  Access Control and Data Governance
    \item Differential Privacy
    \item Data Provenance \& Lineage
    \item Data Integrity
\end{itemize}

The Data Security layer ensures that AI models operate on trustworthy, unaltered data, reducing the likelihood of adversarial manipulation.

\subsubsection{Model Security}

The Model Security Layer is designed to shield AI models from adversarial attacks and emerging cyber threats. AI models are vulnerable to attacks such as adversarial perturbations, model extraction, backdoor injections, and inversion attacks. It is critical to ensure proactive security measures are in place to secure the models used by agentic AI systems. 

The Model Security Layer protects AI models from adversarial inputs, model jailbreaking, data poisoning, and unauthorized access during training, deployment, and operation. Agentic AI systems are becoming increasingly autonomous and interconnected. AI model vulnerabilities can pose significant risks to Agentic operation integrity, confidentiality, and trustworthiness. This layer introduces a set of protections and mitigations for the AI model lifecycle, from training to inference.

The model security layer covers the following controls:
\begin{itemize}
    \item Adversarial Defense Techniques
    \item Model Hardening and Confidentiality
    \item Poisoning and Backdoor Detection
    \item Secure Model Deployment and Execution
\end{itemize}

With these security measures, this layer ensures the integrity, confidentiality and resilience of AI models in adversarial environments.

\subsubsection{Agent Execution and Control}

The Agent Execution and Control Layer is focused on securing the AI agent's operational environment. AI operational environments include AI agents performing their tasks, accessing memory stores, data, and tools to achieve the desired outcome. Given that these agents operate with varying levels of autonomy, it is crucial to regulate their behavior to prevent unauthorized or harmful actions. 

The agent execution and control security layer covers the following controls:
\begin{itemize}
    \item Execution Sandbox 
    \item Policy (behavioral constraints) Enforcement
    \item Runtime \& Safety Verification
    \item Secure API and tools integration controls
\end{itemize}

The agent execution and control layer acts as an agentic AI runtime security and behavioral governor for agentic AI systems. By integrating sandboxing, policy enforcement, runtime validation, and secure integration controls. It ensures that agents can act securely, safely, and predictably in a dynamic environment. These security measures enforce operational discipline while maintaining flexibility for adaptive learning and complex task execution. 

\subsubsection{Accountability \& Trustworthiness}

As agentic AI systems become increasingly autonomous in their decision-making. Hence, accountability and trustworthiness become a key requirements for secure and responsible deployment and use of AI agents. AI Agents must be highly reliable and ethical to build user confidence and mitigate their potential risks.

The accountability and trustworthiness layer covers the following controls:
\begin{itemize}
    \item Explainable AI (XAI) Techniques
    \item Bias Detection and Mitigation
    \item AI System Documentation
    \item  Accountability and Provenance Mechanisms 
    \item Human Oversight and Governance
\end{itemize}

This layer focuses on building accountability, transparency, and interpretability throughout the agentic AI lifecycle. It establishes mechanisms that enable stakeholders, such as developers, auditors, and users, to understand, verify, and trust AI agent behavior. The goal of this layer is to maintain ethical, compliance, and human oversight and ensure that agentic AI systems act within defined ethical and legal boundaries. It ensures that AI agents not only perform securely but also act in a manner that can be justified, audited, and trusted.

\subsubsection{User and Access Management}

The User and Access Management Layer is critical for ensuring identity, authentication, and authorization controls for the agentic AI system. It controls how users, services, and autonomous agents log in, acquire permissions, and interact with the system. It helps manage who can access the system, instruct AI agents, and what they can do, keeping the data and services secure. By using strong security methods, this layer ensures that only approved users and the system can engage with the agentic AI system.

\begin{table*}[h!]
    \caption{MITRE ATLAS Tactics Mapping}
    \label{tab:MITRE ATLAS Mapping}
    \centering
    \begin{tabular}{|p{2.6cm}|p{7.5cm}|p{6.5cm}|} \hline 
         MITRE ATLAS Tactic&  Description& Mapped Security Layer(s)
\\ \hline 
         Reconnaissance&  Gathering information about the target AI system to identify potential vulnerabilities.& Monitoring and Audit Layer\\ \hline 
         Initial Access&  Gaining unauthorized access to the AI system.& User and Access Management, Infrastructure Security Layer\\ \hline 
         Execution&  Running malicious code or commands within the AI system.& Agent Execution and Control Layer\\ \hline 
         Persistence&  Maintaining access to the AI system over time.& Agent Execution and Control, Infrastructure Security Layer\\ \hline 
         Privilege Escalation&  Gaining higher-level permissions within the AI system.& User and Access Management, Infrastructure Security Layer\\ \hline 
         Defense Evasion&  Avoiding detection by security measures within the AI system.& Monitoring and Audit Layer, Model Security Layer
\\ \hline 
         Credential Access&  Stealing or compromising credentials used by the AI system.& User and Access Management Layer
\\ \hline 
         Discovery&  Identifying information about the AI system's environment and configurations.& Monitoring and Audit Layer
\\ \hline 
 Collection& Gathering data from the AI system for exfiltration.&Data Security Layer, Monitoring and Audit Layer
\\ \hline 
 Command and Control& Establishing communication channels to control the AI system remotely.&Agent Execution and Control Layer, Infrastructure Security Layer
\\ \hline 
 Exfiltration& Transferring data from the AI system to an external location.&Data Security Layer, Infrastructure Security Layer
\\ \hline 
         Impact&  Manipulating, disrupting, or destroying the AI system's operations.& Accountability \& Trustworthiness Layer, Agent Execution and Control Layer\\ \hline
    \end{tabular}
\end{table*}

The user and access management layer covers the following controls:

\begin{itemize}
    \item Identity Governance
    \item  Access Control (Authorization)
    \item Multi-Factor Authentication (MFA) and Credential Protection
    \item Privilege Management and Segregation of Duties
    \item Continuous Access Monitoring and Behavioral Analytics 
\end{itemize}

The User and Access Management layer provides the security measure to govern how individuals, services, and autonomous agents authenticate, authorize, and interact with the agentic AI system. 

\subsubsection{Monitoring and Audit Layer}

The Monitoring and Audit Layer gives ongoing visibility into how AI agents behave during operation. It ensures security, compliance, accountability, and monitoring in line with security and governance requirements. Since AI agents operate on their own, continuous monitoring is a must to monitor their interactions, decisions, and activities in real time. It helps to identify unusual behavior, prevent misuse, and ensure compliance with organizational and regulatory standards. This layer sets advanced tracking, behavior analysis, and secure audit methods to protect against outside threats and internal policy violations.

The monitoring and audit layer covers the following controls:
\begin{itemize}
    \item Immutable Logging and Audit Trails 
    \item Continuous Monitoring and Anomaly Detection
    \item   Agent Behavioral Analytics
    \item Threat Intelligence and Adaptive Security Policies
    \item Automated Incident Response Mechanisms
\end{itemize}

These monitoring and audit controls create a detailed monitoring and observability mechanism to protect and ensure the reliability of agentic AI systems and operations. Incorporating continuous monitoring, secure auditing, immutable logging, adaptive policy enforcement, and automatic responses provides assurance, awareness, and protection against the agent AI threats.

\section{Validation}

We conducted a preliminary exercise to map potential threats and assess how well the MAAIS  Framework protects against AI threats and tactics. To evaluate the defensive coverage, we used the MITRE ATLAS (Adversarial Threat Landscape for Artificial-Intelligence Systems) framework for the preliminary validation exercise. MITRE ATLAS \cite{mitre2023atlas} contains a structured taxonomy of known adversarial tactics and techniques targeting AI and machine learning systems. The mapping process helped assess how effectively each security layer mitigates or detects specific categories of adversarial behavior.

This validation process involved analyzing the alignment between the layered framework’s security controls and the threat tactics defined by MITRE ATLAS, such as reconnaissance, model manipulation, privilege escalation, and exfiltration. Each control layer was evaluated for its role in disrupting, detecting, or responding to these tactics. Table \ref{tab:MITRE ATLAS Mapping} shows the mapping of each MITRE ATLAS framework tactics, their description, to each layer within the MAAIS Framework. As seen in the table, this exercise helps confirm that the layered design of the proposed framework effectively addresses the different tactics that attackers might use to compromise agentic AI systems.

\section{Conclusion}

The research contributes to the foundational knowledge of the emerging field of Agentic AI Security by aligning with the MITRE ATLAS framework. The proposed MAAIS Framework adopts a multi-layered security architecture to thoroughly address the security needs of agentic AI systems, mitigate threats, and associated risks. Each layer enforces specific security measures that collectively enhance the confidentiality, integrity, availability, and accountability (CIAA) of AI operations. The framework ensures data security, model robustness, execution control, transparency, access management, and continuous monitoring, establishing a defense-in-depth strategy for AI agents security.  Future work will include mapping it to other regulatory, compliance, and security standards and offering guidance for future standardization or regulatory frameworks. We will also focus on creating clear guidance for each layer of the proposed model that will help researchers and industry practitioners effectively use the layered framework. This will make it easier to integrate the model and improve its application in real-world situations.

\section*{Acknowledgment}

Grammarly was used during the writing process to assist with editing, spell-checking, and grammar improvement. All content and ideas presented in this paper are our own. 

\balance
\printbibliography

\end{document}